\documentstyle[preprint,aps]{revtex}
\tightenlines
\begin{document}
\draft
\title{On the color simple group from chiral electroweak 
       gauge groups}
\author{A. Doff and F. Pisano}
\address{Departamento de F\'\i sica, Universidade Federal do 
         Paran\'a 81531-990 Curitiba PR, Brazil}
\date\today
\maketitle
\begin{abstract}
Following the pioneering Okubo scrutiny of gauge simple groups 
for the quantum chromodynamics we show the constraints coming from 
the wondrous predictive leptoquark-bilepton flavordynamics connecting 
the number of color charges to solution of the flavor question and 
to an electric charge quantization unconstrained from the Dirac, 
Majorana or Dirac--Majorana character of massive neutrino. 
\end{abstract}
\bigskip
\bigskip
Nobody has put to the test the interplay between the 
Fermi-TeV-kTeV~\cite{ktevFnal} 
and the end of space-time Planck scales but everybody believes that there 
is more physics beyond the standard QCD and QFD. 
Since the standard model~\cite{Glashetc} 
of nuclear and electromagnetic interactions is very well 
confirmed~\cite{pdgwww,nu} up to the TeV scale the possibilities 
for chiral gauge semisimple group extensions have not been exhausted yet. 
The most general chiral gauge semisimple group, expanding the number of 
color charges ($n_c$) and the weak isospin group ($m$) is
$$
G_{n_c m_L 1_N} \equiv 
{\rm SU}(n_c)_c\otimes {\rm SU}(m)_L\otimes {\rm U(1)}_N
$$
where the minimal extension is the $G_{331}$ gauge 
symmetry~\cite{pp92,fra92}. 
Although the data accumulated on the scaling violations in deep 
inelastic scattering experiments are consistent with SU(3) gauge 
structure of strong interactions, 
extensions of the color 
sector~\cite{Okubo77} in which quarks transform under the fundamental 
representations of SU($n_c$)$_c$, $n_c = 4,5$ were considered in the 
context of the electroweak standard model preserving the consistency 
at low energies~\cite{Footetal}. 
However, for the first time too many fundamental questions of physics 
are answered and within the minimal semisimple gauge group 
extension of the standard model. 
Although the weak isospin group is minimally enlarged to SU(3)$_L$ 
preserving the color and Abelian group factors there is a first 
unusual capacity in answering fundamental questions only 
by exploring the minimal enlargement of the electroweak gauge group. 
Consider, for instance, the following: 
\begin{enumerate}
\item Each generation is anomalous and the anomalies cancel when 
    the number of leptonic generations is divisible by the number 
    of colors~\cite{pp92,fra92,fp96,DP2000}. There is a relation between 
    the strong and electroweak sectors of the model which does not 
    exist in the standard model with the solution for the 
    flavor question;
\item The electroweak mixing angle, $\theta_{\rm W}$, is limited from 
      above with an upper bound determined by their Landau 
      pole~\cite{Ng94};
\item The neutrino and the charged leptons masses are constrained in the 
cubic seesaw relation
$$
m_{\nu_\ell}\propto\frac{m^3_\ell}{M^2_W}, \quad \ell = e,\mu,\tau
$$
with outcomes for the solar neutrino problem and hot dark 
matter~\cite{fr94}; 
\item The Yukawa couplings have a Peccei--Quinn~\cite{PQ} symmetry 
      which can be extended to all sectors of the Lagrangian with 
      an invisible axion solving the strong-CP problem~\cite{Pal95}; 
\item Spontaneous CP violation in the electroweak sector~\cite{Epele95}. 
      There are several natural sources of explicit and 
      spontaneous CP violations~\cite{vicente};
\item The quark mass hierarchy~\cite{austr};
\item Although the leptoquark-bilepton models do not conserve each 
      generation lepton number $L_\ell$, the neutrinoless double beta 
      decay is forbidden because of the conservation of the quantum 
      number ${\cal F}\equiv L + B$, where $B$ is the barion number 
      and $L=\sum_\ell L_\ell$ is the total lepton number. If this 
      global symmetry is explicitly violated in the Higgs potential, 
      there are 
      contributions to the decay which depend less on the neutrino 
      mass than they do in too many extensions of the standard 
      model~\cite{pt93}. The double beta decay with Majoron emission 
      is possible as well~\cite{fpshe1}; 
\item There is an electric charge quantization without any constraint on 
      the Dirac, Majorana or Dirac--Majorana~\cite{Esposito} character 
      of the massive neutral fermions~\cite{doff}.
\end{enumerate}
\par
Representation contents are determined by embedding the electric 
charge operator
\begin{equation}
\frac{{\cal Q}}{|e|} = (\Lambda_3 + \xi\Lambda_8 + \zeta\Lambda_{15}) + N
\label{opcar}
\end{equation}
in the neutral generators 
$\Lambda_{3,8,15} = \lambda^{\rm SU(4)}_{3,8,15}/2$ of the 
largest 
weak isospin group SU(4) extension and $N$ is the new U(1)$_N$ charge 
equivalent to the electric charge average of the fermions contained 
in each flavor multiplet. If we consider the lightest leptons as the 
fermions  which determine the approximate symmetry, and also independent 
flavor generations, then SU(4)$\times$U(1) is the largest 
non-symmetric gauge group of the electroweak sector. There is no room for 
the chiral semisimple group SU(5)$\times$U(1) if lepton electric charges 
are only $0$, $\pm 1$. The weak hypercharge of the $G_{321}$ standard 
model is
\begin{equation}
\frac{Y}{2} = (\xi\Lambda_8 + \zeta\Lambda_{15})+N
\label{anaa}
\end{equation}
and in the minimal $G_{331}$ leptoquark-bilepton model, $\xi=-\sqrt 3$, 
$\zeta = 0$, are contained 17 gauge vector fields, 
\begin{eqnarray}
{\rm SU}(3)_c & : & g^i_\mu\sim ({\bf 8},{\bf 1},N=0); \quad i=1,2,...,8; 
\nonumber \\
{\rm SU}(3)_L & : & W^j_\mu\sim ({\bf 1},{\bf 8},0); \quad j=1,2,...,8;
\\
{\rm U}(1)_N & : & B_\mu\sim ({\bf 1},{\bf 1},0),
\nonumber
\end{eqnarray}
nine lepton fields connected through charge conjugation of the charged 
fields in three triplets,
\begin{equation} 
L_\ell\sim ({\bf 1},{\bf 3},0), \quad \ell = e, \mu, \tau; 
\label{nonoh}
\end{equation}
three families of quarks, 
\begin{eqnarray}
Q_{1L} & \sim & ({\bf 3},{\bf 3},+2/3) 
\nonumber \\ 
u_R & \sim & ({\bf 3},{\bf 1},+2/3)
\nonumber \\ 
d_R & \sim & ({\bf 3},{\bf 1},-1/3)
\nonumber \\
J_{1R} & \sim & ({\bf 3},{\bf 1},+5/3)
\end{eqnarray}
for the first family, and 
\begin{eqnarray}
Q_{\alpha L} & \sim & ({\bf 3},{\bf\bar 3},-1/3) 
\nonumber \\ 
c_{\alpha R} & \sim & ({\bf 3},{\bf 1},+2/3)
\nonumber \\
s_{\alpha R} & \sim & ({\bf 3},{\bf 1},-1/3)
\nonumber \\
J_{\alpha R} & \sim & ({\bf 3},{\bf 1},-4/3)
\end{eqnarray}
where $\alpha = 2,3$ labels the second and third families. Taking 
into account three color charges we have an amount of 54 quark 
fields. The $J_1$ and $J_\alpha$ leptoquark fermions are color-triplet 
particles with electric charge $\pm\frac{5}{3}$ and $\mp\frac{4}{3}$ 
which carry baryon number and lepton number, 
$B_{J_{1,\alpha}}=+\frac{1}{3}$, and $L_{J_\alpha}=-L_{J_1}=+2$. 
All masses are generated with four multiplets of scalar fields
\begin{eqnarray}
\eta & \sim & ({\bf 1}, {\bf 3}, 0)
\nonumber \\
\rho & \sim & ({\bf 1}, {\bf 3},+1)
\nonumber \\ 
\chi & \sim & ({\bf 1}, {\bf 3},-1)
\nonumber \\ 
S_{ij} & \sim & ({\bf 1},{\bf\bar 6}_S,0)
\end{eqnarray}
and in the symmetric phase of the theory they are parametrized by 30 
real scalar fields. Such unavoidable scalarland is the most 
desirable field sector to have an experimental comprovation, 
since in theories with spontaneous symmetry breaking of the 
gauge symmetry it is essential but the unique field of the standard 
model which does not present evidences is the Higgs scalar boson. 
The total number of massless fields in the $G_{331}$ model 
is 110 and there are not spin-$\frac{3}{2}$ Rarita--Schwinger 
fields. 
\par
Our main purpose is to select the possible color gauge simple groups 
from the leptoquark-bilepton flavordynamics. 
Let us remark that in a theory whith the SU($n_c$)$_c$ gauge simple 
group the 't Hooft~\cite{thft} limit $n_c\rightarrow \infty$ and the 
Maldacena~\cite{Mald} conjecture provide the evidence of the gauge to 
string theories limit. In the four-dimensional super Yang--Mills 
type IIB string theory arises the color confinement and a mass 
gap within the 5-brane of the eleven-dimensional M-theory~\cite{Witten98}. 
The $n_c=3$ standard QCD is an asymptotically free theory including its 
non-perturbative confinement property. The perturbative strong coupling 
constant is
\begin{equation}
\alpha_{\rm s}(q) = \frac{g^2_{{\rm SU}(n_c)_c}(q)}{4\pi} = 4\pi
\left [\beta_0
\ln\left (\frac{q}{\Lambda_{\rm QCD}}\right )^2\right ]^{-1}
\label{strcoupl}
\end{equation}
with
\begin{equation}
\beta_0 = \frac{11}{3} n_c - \frac{2}{3} n_f
\label{eexxuno}
\end{equation}
and the fundamental scale $\Lambda_{\rm QCD}\simeq 250$ 
MeV $\simeq 10^{-3} (\sqrt 2 G_{\rm F})^{-\frac{1}{2}} 
\simeq 246 \times 10^{-3}$ GeV $= 10^{-3}\Lambda_{\rm QFD}$ where 
quarks form the hadrons as a direct effect of the color confinement 
and $n_f$ is the number of quark flavors. The $\Lambda_{\rm QFD}$ is the 
Fermi scale of electroweak spontaneous symmetry breaking 
$G_{321}\rightarrow$ SU(3)$_c\times$U(1)$_{\rm em}$. 
There are two limits, 
\begin{equation}
\lim_{n_c\rightarrow\infty}\alpha_{\rm s} (q) = 0, \quad 
\lim_{q\rightarrow\infty}\alpha_{\rm s} (q) = 0.
\label{limote}
\end{equation}
At high energy, $q^2/\Lambda^2_{\rm QCD}\gg 1$, the strong coupling 
constant is small and the QCD is described by the perturbation theory. 
In 't Hooft original expansion the number of flavors is kept fixed when 
$n_c\rightarrow\infty$. The SU($n_c$)$_c$ exact symmetry is realized 
in the hidden way. 
In the  weak coupling limit
\begin{equation}
a^2\Lambda^2_{\rm QCD} = \exp\left \{-\frac{1}{\beta_0}\left (\frac{4\pi}{ 
g_{{\rm SU}(n_c)_c}(a)}\right )^2 \right \}
\label{acplqcd}
\end{equation}
where $a$ could be the spacing scale of a lattice gauge theory 
of strong couplings the Maldacena conjecture provides a special evidence 
that a string theory comes out from a gauge theory~\cite{DiVe}. 
The $G_{331}$ theory has two anomalies containing the color gauge group. 
Characterizing each triangle anomaly by three generators associated to 
the gauge group they are [SU(3)$_c$]$^3$ and [SU(3)$_c$]$^2$[U(1)$_N$]. 
The pure cubic color anomaly cancel since the QCD has a vector-like 
fermion representation content so there is independent anomaly 
cancellation in each color triplet and the associated antitriplets of 
quarks. Setting the notation for the standard quark chiral flavors
\begin{mathletters}
\begin{eqnarray}
N_{u_R} = N_{c_R} = N_{t_R} & \equiv & N_{U_R}\,,\\ 
\label{xaxii}
N_{d_R} = N_{s_R} = N_{b_R} & \equiv & N_{D_R}\,, 
\label{xqk}
\end{eqnarray}
and
\begin{eqnarray}
N_{Q_{2L}} = N_{Q_{3L}} & \equiv & N_{Q_{\alpha L}}, \\
\label{nqua}
N_{J_{2R}} = N_{J_{3R}} & \equiv & N_{J_{\alpha R}}
\label{nqiu}
\end{eqnarray}
\end{mathletters}
also for the leptoquark flavors, 
the Tr([{\rm SU(3)}$_c$]$^2$[U(1)$_N$]$)=0$ constraint is
\begin{equation}
3(N_{Q_{1L}} + 2N_{Q_{\alpha L}}) - 3(N_{U_R} + N_{D_R}) - 
N_{J_{1R}} - 2 N_{J_{\alpha R}} = 0
\label{ampa}
\end{equation}
and since the $\sum_{L_\ell}N_{L_\ell}$ term vanishes coincides with 
the mixed gravitational-gauge anomaly constraint 
Tr([{\rm graviton}]$^2$[U(1)$_N$])$=0$. Also the 
$N_{Q_{1L}} + 2N_{Q_{\alpha L}}$ term vanishes in the minimal and 
extended leptoquark-bilepton models~\cite{pp92,fra92,sufour}. 
\par
Being ${\rm N}_\ell$ and ${\rm N}_{\rm q}$ the number of lepton 
and quark generations 
let us consider the SU($n_c$)$_c$ possibilities for $n_c\geq 3$ where 
the ${\rm N}_\ell = {\rm N}_{\rm q} = {\rm N}_{\rm generations}$ 
coincidence is evaded. Denoting as $n_{\bf m}$ and 
$n_{{\bar{\bf m}}}$ the number of quark generation multiplets 
transforming 
as ${\bf m}$ and ${{\bar{\bf m}}}$ in the fundamental representation 
under the SU($m$)$_L$ flavor group factor we have the universality 
breaking condition in the lepton sector
\begin{mathletters}
\begin{equation}
{\rm N}_\ell = |n_c\,(n_{\bf m} - n_{{\bar{\bf m}}})|,
\label{gng}
\end{equation}
and 
\begin{equation}
{\rm N}_{\rm q} = n_{\bf m} + n_{\bar{\bf m}}
\label{unaltr}
\end{equation}
\end{mathletters}
for the number of quark flavor generations. The condition in Eq.~(\ref{gng}) 
involves the following possibilities: (1) $n_{\bf m} > n_{\bar{\bf m}}$, 
when the leptons must transform as ${\bar{\bf m}}$; 
(2) $n_{\bf m} < n_{\bar{\bf m}}$ when the lepton multiplets are 
attributed to the ${\bf m}$ representation; 
(3) $n_{\bf m} = n_{\bar{\bf m}}$. 
For $n_{\bf m} > n_{\bar{\bf m}}$ and in the case of even $n_c$, 
$n_c = 2k$, $k\geq 2$ we have the ratio
\begin{equation}
\frac{n_{\bf m}}{n_{\bar{\bf m}}} = 
\frac{2k + 1}{2k - 1}
\label{ratgen}
\end{equation}
but for odd $n_c = 2k + 1$ the ratio is
\begin{equation}
\frac{n_{\bf m}}{n_{\bar{\bf m}}} = 1+\frac{1}{k}.
\label{impa}
\end{equation}
The SU(5)$_c$ group consistent with the standard 
flavordynamics~\cite{Footetal} satisfies the ${\rm N}_\ell = 
{\rm N}_{\rm q}$ condition if $n_{\bf m}/n_{\bar{\bf m}} = 3/2$ for 
$k=2$ with five generations in a universal representation content. 
The ratios of the lepton and quark generations number with the number of 
color charges are
\begin{equation}
\frac{{\rm N}_\ell}{n_c} = \frac{{\rm N_{\rm q}}}{n_c} = 
\frac{{\rm N}_{\rm generations}}{n_c} = k,
\label{fin}
\end{equation}
for all $k\in\{1,2,3,...\}$ so as to 
\begin{equation}
\lim_{{n_c}\rightarrow\infty}\frac{{\rm N}_{\rm generations}}{n_c} = 0
\label{fndu}
\end{equation}
for a finite $k$. 
\par
When $k\rightarrow\infty$ then N$_{\rm generations}\rightarrow\infty$ 
for $n_c=3$ or $n_c=4,5$~\cite{Footetal} but when $n_c\rightarrow\infty$ 
the $\frac{\infty}{\infty}$ indetermination arises. This could be seen as 
an intrinsic limitation of the theory with conformal invariance due to the 
horizontal replication of fundamental matter fields pointing from the 
particle to the string elementarity level.


\begin{references}
\bibitem{ktevFnal} {\sf http://
                   fnphyx-www.fnal.gov/experiments/ktev/epsprime.html}.
\bibitem{Glashetc} S. L. Glashow, Nucl. Phys. {\bf 22}, 279 (1961); 
                   S. Weinberg, Phys. Rev. Lett. {\bf 19}, 1264 (1967); 
                   A. Salam, in {\it Elementary Particle Theory}, 
                   edited by N. Svartholm (Almquist and Wiksell, 
                   Stockholm, 1968), p. 367; 
                   S. L. Glashow, J. Iliopoulos and L. Maiani, Phys. 
                   Rev. D {\bf 2}, 1285 (1970); 
                   N. Cabibbo, Phys. Rev. Lett. {\bf 10}, 531 (1963); 
                   M. Kobayashi and K. Maskawa, Prog. Theor. Phys. {\bf 49}, 
                   652 (1973); O. W. Greenberg, Phys. Rev. Lett. {\bf 13}, 
                   598 (1964); M. Gell-Mann, Acta Phys. Austriaca, Suppl. 
                   {\bf IX}, 733 (1972); H. Fritzsch, M. Gell-Mann and 
                   H. Leutwyler, Phys. Lett. {\bf 47}, 365 (1973); 
                   S. Weinberg, Phys. Rev. Lett. {\bf 31}, 494 (1973); 
                   D. Gross and F. Wilczek, Phys. Rev. Lett. {\bf 30}, 
                   1343 (1973).
\bibitem{pdgwww} Particle Data Group, D. E. Groom {\it et al.}, Review 
                 of Particle Physics, Eur. Phys. J. C {\bf 15}, 1-878 
                 (2000), {\sf http://pdg.lbl.gov}.
\bibitem{nu} Evidence of neutrino flavor oscillation in atmospheric, solar 
             and accelerator data were reported since 1998, 
             Y. Fukuda {\it et al.}, (Super-Kamiokande collab.), Phys. 
             Lett. B {\bf 433}, 9 (1998); B {\bf 436}, 33 (1998); 
             Phys. Rev. Lett. {\bf 81}, 1158, 1562 (1998); 
             {\bf 82}, 2430, 2624 (1999); 
             C. Athanassopoulos {\it et al.}, (LSND collab.), 
             Phys. Rev. Lett. {\bf 81}, 1774 (1998).
\bibitem{pp92} F. Pisano and V. Pleitez, ``Neutrinoless double beta decay 
               and doubly charged gauge bosons'' IFT-P-017-91 (Aug 1991), 
               {\sf www.ift.unesp.br}; 
               F. Pisano and V. Pleitez, Phys. Rev. D {\bf 46}, 410 (1992); 
               R. Foot {\it et al.}, Phys. Rev. D {\bf 47}, 4158 (1993).
\bibitem{fra92} P. H. Frampton, Phys. Rev. Lett. {\bf 69}, 2889 (1992).
\bibitem{Okubo77} Susumu Okubo, Phys. Rev. D {\bf 16}, 3528, 3535 (1977).
\bibitem{Footetal} R. Foot, Phys. Rev. D {\bf 40}, 3136 (1989); 
                   R. Foot and O. F. Hern\'andez, Phys. Rev. D {\bf 41}, 
                   2283 (1990); Phys. Rev. D {\bf 42}, 948 (1990); 
                   R. Foot, O. Hern\'andez and T. G. Rizzo, 
                   Phys. Lett. B {\bf 246}, 183 (1990); B {\bf 261}, 
                   153 (1991).
\bibitem{fp96} F. Pisano, Mod. Phys. Lett. A {\bf 11}, 2639 (1996).
\bibitem{DP2000} A. Doff and F. Pisano, Mod. Phys. Lett. A {\bf 15}, 1471 
                 (2000).
\bibitem{Ng94} D. Ng, Phys. Rev. D {\bf 49}, 4805 (1994).
\bibitem{fr94} P. H. Frampton, P. I. Krastev and J. T. Liu, 
               Mod. Phys. Lett. A {\bf 9}, 761 (1994).
\bibitem{PQ} R. D. Peccei and H. Quinn, Phys. Rev. Lett. {\bf 38}, 1440 
             (1977); Phys. Rev. D {\bf 16}, 1791 (1977);.
\bibitem{Pal95} Palash B. Pal, Phys. Rev. D {\bf 52}, 1659 (1995).
\bibitem{Epele95} L. Epele, H. Fanchiotti, C. Garc\'\i a-Canal and 
                  D. G\'omez Dumm, Phys. Lett. B {\bf 343}, 291 (1995).  
\bibitem{vicente} J. C. Montero, V. Pleitez and O. Ravinez, 
                  Phys. Rev. D {\bf 60}, 076003 (1999); 
                  J. C. Montero, C. Pires and V. Pleitez, Phys. Rev. 
                  D {\bf 60}, 115003 (1999).
\bibitem{austr} M. B. Tully and G. C. Joshi, Mod. Phys. Lett. A {\bf 13}, 
                2065 (1998); P. H. Frampton and Otto C. W. Kong, 
                Phys. Rev. D {\bf 55}, 5501 (1997).
\bibitem{pt93} V. Pleitez and M. D. Tonasse, Phys. Rev. D {\bf 48}, 
               5274 (1993).
\bibitem{fpshe1} F. Pisano and S. Shelly Sharma, Phys. Rev. D {\bf 57}, 
                 5670 (1998).
\bibitem{Esposito} S. Esposito and G. Capone, Zeits. f\"ur Physik 
                   C {\bf 70}, 55 (1996), S. Esposito and N. Tancredi, 
                   Mod. Phys. Lett. A {\bf 12}, 1829 (1997); Eur. Phys. J. 
                   C {\bf 4}, 221 (1998); S. Esposito, Int. J. Mod. Phys. 
                   A {\bf 13}, 5023 (1998).
\bibitem{doff} A. Doff and F. Pisano, Mod. Phys. Lett. A {\bf 14}, 1133 
               (1999).
\bibitem{thft} G. 't Hooft, Nucl. Phys. B {\bf 72}, 461 (1974).
\bibitem{Mald} J. Maldacena, Adv. Theor. Math. Phys. {\bf 2}, 231 (1998).
\bibitem{Witten98} E. Witten, Adv. Theor. Math. Phys. {\bf 2}, 505 (1998).
\bibitem{DiVe} P. Di Vecchia, Acta Physica Austriaca, Suppl. {\bf XXII}, 
               341 (1980); Phys. Lett. B {\bf 85}, 357 (1979); P. Di Vecchia 
               and G. Veneziano, Nucl. Phys. B {\bf 171}, 253 (1980).
\bibitem{sufour} F. Pisano and T. A. Tran, {\it Anomaly cancellation in a 
                 class of chiral flavor gauge models}, ICTP Report IC/93/200, 
                 Proc. of the 14th National Meeting on Particles and Fields, 
                 The Brazilian Phys. Soc., p. 322 (1993); 
                 R. Foot, H. N. Long and T. A. Tran, Phys. Rev. D {\bf 50}, 
                 R34 (1994); F. Pisano and V. Pleitez, Phys. Rev. D 
                 {\bf 51}, 3865 (1995).
\end{references}
\end{document}